%% file: oup-authoring-template.tex
\documentclass[namedate,webpdf,modern,small]{oup-authoring-template}

\onecolumn 

\usepackage{booktabs}
\usepackage{setspace}
\usepackage{graphicx}
\usepackage{amsmath, amssymb}
\usepackage{microtype}
\usepackage{hyperref}
\usepackage{caption}
\usepackage{tabularx}
\usepackage{array}

\usepackage{ragged2e}
\newcolumntype{Y}{>{\RaggedRight\arraybackslash}X}

\usepackage[most]{tcolorbox}

\tcbset{
  mspibox/.style={
    enhanced,
    boxrule=0pt,
    frame hidden,
    colback=black!4,
    arc=2mm,
    left=2mm,right=2mm,top=1.5mm,bottom=1.5mm,
    before skip=10pt, after skip=10pt
  }
}

\newcolumntype{Y}{>{\raggedright\arraybackslash}X}

\newcommand{\indic}{\mathbb{I}}
\newcommand{\R}{\mathbb{R}}

\graphicspath{{Fig/}}

\theoremstyle{thmstyleone}%
\theoremstyle{thmstyletwo}%
\theoremstyle{thmstylethree}%

\usepackage{fancyhdr}
\begin{document}

\journaltitle{}
\DOI{}
\copyrightyear{}
\pubyear{}
\vol{}
\issue{}
\access{}
\appnotes{}



\title[Algorithmic Monitoring]{Algorithmic Monitoring: Measuring Market Stress with Machine Learning}


\author[]{Marc Schmitt\ORCID{0000-0003-4550-2963}}


\address[]{\orgdiv{Department of Computer Science}, \orgname{University of Oxford}}


\received{Date}{0}{Year}
\revised{Date}{0}{Year}
\accepted{Date}{0}{Year}



\abstract{I construct a Market Stress Probability Index (MSPI) that estimates the probability of high stress in the U.S. equity market one month ahead using information from the cross-section of individual stocks. Using CRSP daily data, each month is summarized by a set of interpretable cross-sectional fragility signals and mapped into a forward-looking stress probability via an L1-regularized logistic regression in a real-time expanding-window design. Out of sample, MSPI tracks major stress episodes and improves discrimination and accuracy relative to a parsimonious benchmark based on lagged market return and realized volatility, delivering calibrated stress probabilities on an economically meaningful scale. Further, I illustrate how MSPI can be used as a probability-based measurement object in financial econometrics. The resulting index provides a transparent and easily updated measure of near-term equity-market stress risk.}



\maketitle
\thispagestyle{fancy}


\vspace{2em}

Financial stress is a latent state of the economy and the financial system. It is observed primarily through its manifestations in asset prices, volatility, and market functioning. Because stress is not directly measurable, empirical macro-finance typically relies on observable proxies or composite indices. Although widely used, such measures can be difficult to interpret, may rely on data outside equity markets, or may not be available at high frequency or consistently over long historical spans.

Over the past two decades, the informational and institutional environment in which financial stress arises has changed substantially. Trading, liquidity provision, and order handling have become increasingly electronic and algorithmically mediated. A large empirical literature shows that algorithmic trading meaningfully affects market quality and liquidity provision in normal times \citep{HendershottJonesMenkveld2011}, while extreme episodes highlight that the same automation can interact with feedback effects under stress \citep{KirilenkoKyleSamadiTuzun2017}. Related evidence from foreign exchange markets similarly finds that algorithmic trading affects liquidity provision and price efficiency \citep{CHABOUD2014}. More broadly, changes in trading venue design are also a defining feature of modern algorithmic markets; for example, \cite{HalimRiyantoRoyWang2025} study how dark trading affects financial markets and document adverse effects on market quality. In such an environment, empirical macro-finance increasingly benefits from stress measures that are (i) computed from high-frequency market data available in real time, (ii) robust to rapidly changing trading microstructure, and (iii) easily and transparently updated as new data arrive. Recent evidence further shows that the behavior of trading algorithms in electronic markets can itself be modeled and predicted using statistical methods, reinforcing the view that monitoring in modern financial markets must be algorithmic in nature \citep{cartea2025statistical}.

I propose a transparent and fully reproducible equity-only measure: the Market Stress Probability Index (MSPI). The key idea is that stress leaves a distinctive footprint in the cross-section of individual stock returns and trading activity. When market conditions deteriorate, return dispersion increases, downside extremes become more frequent, and cross-sectional higher moments shift in ways that reflect elevated tail risk and co-movement. I summarize these cross-sectional patterns into a set of monthly fragility signals and use a machine learning classifier to translate them into a one-month-ahead probability of entering a stress state.

A practical implication of today’s electronic market design is that monitoring itself must be algorithmic. When trading, liquidity provision, and risk controls are increasingly automated, market-wide stress can be triggered or amplified by fast feedback mechanisms—such as synchronized liquidity withdrawal, correlated deleveraging, or order-handling frictions—that are difficult to diagnose in real time using discretionary oversight alone. This shifts a classic financial-econometric problem—the measurement of latent stress states—into a setting in which the input data are high-dimensional and the monitoring output must be continuously updated, reproducible, and auditable. This paper frames MSPI as such an algorithmic monitoring object: a parsimonious, reproducible probability signal designed for real-time use in modern electronic markets \citep{HendershottJonesMenkveld2011,KirilenkoKyleSamadiTuzun2017,CHABOUD2014}.

MSPI is designed as a financial-econometric measurement pipeline for this environment: it compresses cross-sectional fragility in returns and trading activity into a single, calibrated probability that the next month will be a stress regime. The probability scale is operationally convenient, as it supports transparent threshold rules and monitoring escalation protocols, and econometrically convenient, as it delivers a state variable that can be used directly in standard empirical designs.

While the baseline MSPI mapping is intentionally sparse and interpretable (lasso logit), I also compare it to more flexible nonlinear machine-learning classifiers that are widely used in financial econometrics, including random forests and gradient boosting. This horse race serves two purposes. First, it verifies that the empirical gains are not an artifact of a particular functional form. Second, it clarifies the transparency–flexibility trade-off in a setting where the object of interest is a real-time probability measure rather than a purely predictive black box. The main result is that the parsimonious MSPI specification remains highly competitive out of sample: even when flexible nonparametric learners are tuned and evaluated within the same expanding-window design, they do not consistently dominate MSPI in terms of both discrimination and probability accuracy.

My contribution is threefold.
First, I introduce algorithmic monitoring as a financial-econometric measurement task and construct a transparent, equity-only system that outputs a real-time probability of entering a stress regime over the next month. The construction is strictly real time: predictors and labels are formed using only information available at the time of prediction, and the mapping is re-estimated in an expanding-window design. This makes MSPI suitable as a monitoring input rather than an ex post descriptive index.
Second, I connect this measurement object to modern predictive econometrics. I treat the baseline lasso-logit as an interpretable statistical-learning mapping and compare it to nonlinear machine-learning methods (random forests and gradient-boosted trees) under the same information set and the same real-time protocol. This comparison isolates whether additional flexibility improves out-of-sample discrimination, probability accuracy, and calibration beyond a sparse, communicable classifier.
Third, I provide an applied interpretation that is natural in financial econometrics. MSPI levels and MSPI innovations summarize near-term stress risk on a calibrated probability scale and are informative about subsequent risk outcomes (e.g., realized market volatility), constructed from the same CRSP universe under the same timing discipline.
Throughout, statistical learning is treated as a measurement tool rather than as a black-box forecasting device. The learning step aggregates multiple noisy cross-sectional fragility signals into a single forward-looking probability-of-stress state variable that is transparent, reproducible, and implementable in real time.

\begin{tcolorbox}[
  enhanced,
  breakable,
  title=\textbf{Box 1. Algorithmic Monitoring in Algorithmic Markets},
  colback=white,
  colframe=black!80,
  fonttitle=\bfseries
]

\noindent \textbf{Why monitoring has changed.}
Modern equity markets are electronic and algorithmically mediated: order placement, execution, and liquidity provision are automated, adaptive, and high-frequency. In such environments, market fragility can arise from fast feedback loops and endogenous crowding into execution or risk-control rules. Stress need not originate solely from macroeconomic or firm-level fundamentals; it can also be amplified by market microstructure and algorithmic interaction.

\medskip
\noindent \textbf{Why the monitor must be algorithmic.}
When the objects being monitored are themselves algorithmic, the monitoring layer faces an asymmetry of speed and scale. Purely manual surveillance struggles to aggregate the many weak, noisy signals generated across thousands of securities and trading days in a timely and consistent manner. This motivates algorithmic monitoring: the use of transparent statistical-learning systems to compress high-dimensional market information into interpretable, real-time state variables that can be tracked, backtested, and communicated.

\medskip
\noindent \textbf{What MSPI provides.}
MSPI operationalizes this idea by mapping cross-sectional fragility signals constructed from CRSP data into a one-month-ahead probability of entering a stress regime. The probability scale supports protocol-based interpretation (e.g., heightened monitoring or scenario checks) and delivers a state variable that is comparable across time and estimation windows.

\end{tcolorbox}

The remainder of the paper proceeds as follows. Section~\ref{sec:background} situates MSPI in the literatures on financial-stress measurement and probabilistic forecasting with statistical learning. Section~\ref{sec:data} describes the CRSP data and sample construction. Section~\ref{sec:signals} defines the cross-sectional fragility signals. Section~\ref{sec:mspi_method} defines the stress label and the real-time learning protocol used to construct MSPI. Section~\ref{sec:results} reports out-of-sample results. Section~\ref{sec:econ} provides a financial interpretation, including realized outcomes across MSPI probability bins and the construction of stress-risk innovations. Section~\ref{sec:conclusion} concludes.

\medskip

\section{Background}
\label{sec:background}

Recent advances in machine learning have expanded the feasibility of extracting stable signals from large, noisy, and high-dimensional datasets \citep{LeCun2015}. From an economic perspective, accumulating evidence suggests that machine learning technologies are likely to have general-purpose characteristics with broad impacts across sectors \citep{Goldfarb2023}. In financial markets, this shifts the frontier of feasible measurement. Rather than relying on a single noisy proxy, researchers can use disciplined learning procedures to aggregate multiple weak signals into transparent, real-time state variables.

\subsection{Financial stress and measurement objectives}
Financial stress indices are commonly used as state variables summarizing disruptions in market functioning, risk appetite, and funding conditions. An important conceptual distinction is between financial conditions (broad, often persistent variation in the ease of financing) and financial stress (acute, nonlinear episodes associated with spikes in uncertainty, volatility, and market dislocations). A large empirical literature develops composite indicators that summarize stress from multiple markets, often with the goal of monitoring and macro-finance applications (e.g., stress episodes and downturns in international data). Related work studies spillovers in extreme downside risk and volatility across markets, highlighting that financial stress can propagate through tail-dependent dynamics and is empirically relevant beyond any single market segment \citep{HONG2009271,zhang2021stock}.

A practical challenge is that stress is latent and measurement is noisy. Existing indices vary in information sets, aggregation schemes, and whether they target (i) an ex post stress level, (ii) a real-time ``nowcast'' of current conditions, or (iii) a forward-looking probability of entering a stress regime. For example, some approaches build stress measures from broad financial indicators and document their relationship with downturns and recoveries (e.g., \citealp{CARDARELLI201178}). Others construct financial stress indices for specific economies and evaluate their time-series behavior around episodes (e.g., \citealp{IllingLiu2006}). The MSPI contribution is to target a probability-of-stress object one month ahead using an equity-only, transparent feature set. This focus on probability outputs motivates an explicit emphasis on probability accuracy and reliability (calibration), not just ranking performance.

Related work in macroeconomics emphasises that constructing transparent indices can substantially improve empirical measurement of latent states, such as uncertainty \citep{BakerBloomDavis2016}. A complementary approach measures downside states directly by focusing on tail risks in macroeconomic outcomes; \cite{LoriaMatthesZhang2025} propose a time-varying measure of macroeconomic tail risk that is informative about macro-financial conditions. In financial economics, a closely related literature develops market-based measures of systemic risk and stress that formalize tail dependence, downside exposure, and capital shortfalls, including CoVaR \citep{AdrianBrunnermeier2016}, capital shortfall measures \citep{AcharyaEngleRichardson2012}, and conditional capital shortfall metrics such as SRISK \citep{BrownleesEngle2017}. These approaches demonstrate the value of explicitly modeling tail events and stress states using market data. The equity-only design adopted here complements this literature by focusing on the cross-section of equity returns and trading activity to construct a forward-looking probability of a near-term stress regime that is directly implementable in real time.

A large related literature also constructs financial stress and financial conditions indices by combining information from multiple markets and institutions, often using composites of spreads, volatilities, and funding conditions. Prominent examples include the Kansas City Financial Stress Index \citep{HakkioKeeton2009}, the Chicago Fed National Financial Conditions Index \citep{BraveButters2011}, and the ECB Composite Indicator of Systemic Stress (CISS) \citep{HolloKremerLoDuca2012}. Relative to such multi-market composites, the equity-only design here isolates stress information embedded in a single, widely available dataset (CRSP) and produces an explicitly forward-looking probability measure that is straightforward to update and backtest.

\subsection{Equity-market based indicators}
Many widely used market-based risk indicators rely on derivatives markets (for example, volatility indices). A long literature uses option-implied volatility as a forward-looking measure of risk and uncertainty (e.g., \citealp{Whaley1993}). In contrast, MSPI is designed to be equity-only and long-span: it uses only individual-stock returns and trading variables, which are broadly available historically and in real time.

A complementary strand of work shows that realized volatility constructed from high-frequency return observations (at daily or intra-daily frequency) provides a powerful empirical measure of time-varying risk (e.g., \citealp{AndersenBollerslevDieboldLabys2003}; \citealp{BarndorffNielsenShephard2002}). MSPI borrows this ``measurement-first'' logic, but targets a different object: rather than measuring realized risk contemporaneously, it maps cross-sectional fragility signals into a forward-looking probability of entering a stress regime in the near future. The equity-only design also facilitates transparent backtesting and disciplined real-time implementation.

\subsection{Statistical learning as measurement}
Finally, the paper connects to the growing literature using machine learning for economic and financial measurement. \cite{MullainathanSpiess2017} emphasise how prediction tools can be deployed as part of an applied econometric workflow, while \cite{GuKellyXiu2020} illustrate how regularization and machine learning can improve measurement of key financial objects such as risk premia. In contrast to high-dimensional black-box implementations, the baseline MSPI specification uses a parsimonious and transparent lasso-logit mapping that is easy to replicate, communicate, and deploy in real time.

While deep neural networks have achieved remarkable success in domains with strong structural regularities such as images, speech, and text \citep{LeCun2015}, their advantages are less clear in tabular economic and financial datasets characterized by low signal-to-noise ratios, limited effective sample sizes, and unstable predictive relationships \citep{GuKellyXiu2020,Schmitt2023Deep,10.1093/jjfinec/nbae028}. Accordingly, this paper focuses on sparse generalized linear models as the baseline specification and uses random forests and gradient-boosted trees as representative nonlinear alternatives. This modeling choice reflects the classical econometric concern of balancing flexible fit against parsimonious structure in finite samples \citep{hardle1993comparing}.

From a predictive-econometrics perspective, the baseline MSPI specification uses sparsity to control overfitting and enhance interpretability. L1 regularization (\citealp{Tibshirani1996}) is attractive in monthly macro-finance settings with limited effective sample sizes because it encourages parsimonious models and reduces estimation variance. At the same time, stress dynamics may involve nonlinearities and interactions. To assess whether added flexibility improves out-of-sample performance, the paper benchmarks the sparse lasso-logit mapping against two widely used nonlinear learners: random forests (\citealp{Breiman2001}) and gradient-boosted trees (\citealp{Friedman2001}). Importantly, all models are evaluated under the same real-time expanding-window design with fixed hyperparameters after initial tuning to avoid implicit look-ahead.

\subsection{Probability forecasting and calibration}
Because the object of interest is a probability-of-stress, evaluation must go beyond rank-based discrimination metrics. Proper scoring rules provide a coherent way to evaluate probabilistic forecasts. The Brier score (\citealp{Brier1950}) and log loss (negative log-likelihood) reward sharp but accurate probabilities and penalize overconfident errors. More generally, strictly proper scoring rules formalize why probability accuracy is central when forecasts are interpreted as beliefs over states (\citealp{GneitingRaftery2007}).

A related issue is calibration: well-calibrated probabilities match empirical frequencies in the appropriate conditioning sets. In practice, flexible classifiers can deliver strong ranking performance while producing distorted probability levels. A standard remedy is to apply an explicit calibration map estimated on held-out data or within each training window, such as Platt scaling (\citealp{Platt1999}) or related score-to-probability transformations (\citealp{ZadroznyElkan2002}). In this paper, calibration is treated as part of the real-time forecasting pipeline for the nonlinear benchmarks, and probability metrics are reported on the calibrated scale.

A probability output is economically meaningful because it can be mapped into actions via threshold rules and expected-loss calculations. This makes forecast evaluation fundamentally different from pure classification: the goal is not only to rank stress months, but to produce probabilities that are accurate and reliable on an absolute scale. Proper scoring rules therefore provide a principled foundation for model comparison in this setting \citep{Brier1950,GneitingRaftery2007}, and explicit calibration is a natural component of the forecasting pipeline for flexible learners \citep{Platt1999,ZadroznyElkan2002}.


\section{Data}\label{sec:data}

This study uses CRSP daily stock data accessed via WRDS for U.S. common stocks. The main inputs are individual-stock daily returns and trading variables (price and volume). To focus on standard common equity claims, the analysis applies conventional filters: (i) ordinary common shares (CRSP share codes 10 and 11), (ii) NYSE/AMEX/NASDAQ listings (exchange codes 1, 2, and 3), (iii) non-missing returns and prices, and (iv) absolute price at least \$1. Daily observations are linked to CRSP security identifiers using CRSP name-history information to apply these filters consistently over time.

The study also uses the CRSP daily value-weighted market index return to construct aggregate market variables. Monthly market returns are obtained by compounding daily index returns within each month. Monthly realized market volatility is computed as the within-month standard deviation of daily value-weighted returns, annualized by $\sqrt{252}$.

Cross-sectional predictors are constructed at the monthly frequency by computing daily cross-sectional statistics across eligible stocks and averaging within each month (e.g., cross-sectional moments, extreme-return shares, and trading-intensity proxies). The modeling sample begins once sufficient history exists to compute the expanding real-time volatility quantile used in the stress definition, and out-of-sample evaluation begins after an initial 120-month training window (Section~\ref{sec:mspi_method}).

\section{Constructing Cross-Sectional Fragility Signals}\label{sec:signals}

Let $r_{i,d}$ denote the daily return of stock $i$ on trading day $d$, and let $N_d$ be the number of stocks available on day $d$. Define the daily cross-sectional mean return
\begin{equation}
\bar{r}_d \equiv \frac{1}{N_d}\sum_{i=1}^{N_d} r_{i,d}.
\end{equation}
Daily cross-sectional dispersion and higher moments are computed as:
\begin{align}
\sigma^{xs}_d &\equiv \sqrt{\frac{1}{N_d}\sum_{i=1}^{N_d}\left(r_{i,d}-\bar{r}_d\right)^2}, \\
\text{Skew}^{xs}_d &\equiv \frac{\frac{1}{N_d}\sum_{i=1}^{N_d}\left(r_{i,d}-\bar{r}_d\right)^3}{\left(\sigma^{xs}_d\right)^3}, \\
\text{Kurt}^{xs}_d &\equiv \frac{\frac{1}{N_d}\sum_{i=1}^{N_d}\left(r_{i,d}-\bar{r}_d\right)^4}{\left(\sigma^{xs}_d\right)^4}.
\end{align}
To capture tail-like behavior in a robust way, the fraction of extreme winners and losers is computed as:
\begin{align}
\text{Frac}^{dn}_d(\tau) &\equiv \frac{1}{N_d}\sum_{i=1}^{N_d}\indic\left\{ r_{i,d}\le -\tau \right\},\\
\text{Frac}^{up}_d(\tau) &\equiv \frac{1}{N_d}\sum_{i=1}^{N_d}\indic\left\{ r_{i,d}\ge \tau \right\},
\end{align}
where $\tau=5\%$ in the baseline implementation. Finally, the analysis includes trading-intensity proxies computed as daily cross-sectional averages: average log volume, average dollar volume, and average turnover (volume scaled by shares outstanding).

Monthly features are constructed by aggregating daily measures within month. Specifically, for any daily statistic $z_d$, the month-$t$ feature is defined as
\begin{equation}
Z_t \equiv \frac{1}{D_t}\sum_{d\in t} z_d,
\end{equation}
where $D_t$ is the number of trading days in month $t$.

These cross-sectional moments are motivated by a simple fragility intuition. In tranquil periods, daily return variation is relatively well diversified across the cross-section and extreme outcomes are rare. In contrast, stress is often characterized by (i) elevated dispersion, reflecting heterogeneous repricing and breakdowns in correlation structure; (ii) a higher share of extreme negative returns, reflecting clustered downside tail events; and (iii) shifts in cross-sectional skewness and kurtosis, reflecting asymmetry and fat tails in realized outcomes. Importantly, tail risk can be estimated directly from the cross-section of equity returns without relying on option-implied measures, which can be valuable when researchers require long-span, equity-only data. For example, \citet{KellyJiang2014} propose a cross-section-based tail risk measure that varies over time and is informative about macro-financial conditions. The feature design in this paper follows the same general principle—extracting latent stress-relevant information from the cross-section—but targets a different measurement object: a real-time probability of a near-term stress regime.

Table~\ref{tab:features} summarizes the final feature set used to construct MSPI. These features are designed to be interpretable and to reflect intuitive channels through which stress affects equity markets: dispersion (fragmentation of returns), tails (extreme outcomes), higher moments (asymmetry and fat tails), and trading intensity (market participation and turnover).

\begin{table*}[!htbp]
\centering
\caption{Cross-sectional fragility signals (monthly predictors)}
\label{tab:features}
\small
\setlength{\tabcolsep}{6pt}
\renewcommand{\arraystretch}{1.05}
\begin{tabularx}{\textwidth}{@{}l Y Y@{}}
\toprule
\textbf{Feature (month $t$)} 
& \textbf{Definition (computed daily and averaged within month)} 
& \textbf{Interpretation / role} \\
\midrule
$n_{\text{stocks}}$ 
& Average number of eligible stocks $N_d$ in month $t$. 
& Coverage / breadth; eligibility changes. \\

$\sigma^{xs}$ 
& Average cross-sectional standard deviation of daily returns, $\sigma^{xs}_d$. 
& Dispersion; heterogeneous shocks / fragmentation. \\

$\text{Skew}^{xs}$ 
& Average cross-sectional skewness of daily returns, $\text{Skew}^{xs}_d$. 
& Asymmetry; concentrated downside. \\

$\text{Kurt}^{xs}$ 
& Average cross-sectional kurtosis of daily returns, $\text{Kurt}^{xs}_d$. 
& Tail thickness; extreme cross-sectional moves. \\

$\overline{|r|}$ 
& Average cross-sectional mean absolute return, $\frac{1}{N_d}\sum_i |r_{i,d}|$. 
& Amplitude of price moves. \\

$\text{Frac}^{dn}(5\%)$ 
& Average fraction of stocks with daily return $\le -5\%$. 
& Breadth of large negative moves. \\

$\text{Frac}^{up}(5\%)$ 
& Average fraction of stocks with daily return $\ge +5\%$. 
& Breadth of large positive moves. \\

$\overline{\log(1+\text{Vol})}$ 
& Average of $\log(1+\text{volume})$ across stocks. 
& Trading intensity; activity spikes. \\

$\overline{\$Vol}$ 
& Average dollar volume, $|P_{i,d}|\times \text{Vol}_{i,d}$. 
& Liquidity usage; dollars traded. \\

$\overline{\text{Turn}}$ 
& Average turnover, $\text{Vol}_{i,d}/\text{SharesOut}_{i,d}$. 
& Turnover / churn; rapid reallocation. \\
\bottomrule
\end{tabularx}

\vspace{0.3em}
\begin{minipage}{\textwidth}
\footnotesize
\emph{Note:} The signals fall into three broad groups: cross-sectional distributional moments (dispersion, skewness, kurtosis), tail participation measures (Frac$^{dn}$, Frac$^{up}$), and trading-intensity proxies (volume, dollar volume, turnover). In aggregate, they summarize breadth, asymmetry, and liquidity usage in a form suitable for real-time market-stress monitoring.
\end{minipage}

\end{table*}

\section{Defining Stress and Learning the Stress Probability Index}\label{sec:mspi_method}

\subsection{Stress definition}

Let $R^{mkt}_t$ denote the CRSP value-weighted market return in month $t$, and let $\sigma^{mkt}_t$ denote the corresponding realized market volatility (annualized) constructed from daily value-weighted returns within month $t$.

A month is defined as a stress month if either market returns are sharply negative or realized volatility is unusually high relative to recent history. Formally,
\begin{equation}
S_t \equiv \indic\left\{ R^{mkt}_t \le c_R \right\} \ \ \vee \ \ \indic\left\{ \sigma^{mkt}_t \ge q_{t-1}(\alpha) \right\},
\label{eq:stress_def}
\end{equation}
where $c_R=-5\%$ is a fixed downside threshold and $q_{t-1}(\alpha)$ is the \emph{expanding, real-time} $\alpha$-quantile of realized volatility computed using data available up to month $t-1$. In the baseline, $\alpha=0.90$. Using $q_{t-1}(\alpha)$ ensures that stress labeling is implementable in real time and does not use information from the future.

The stress definition combines two complementary manifestations of stress. Large negative monthly returns capture crash-like episodes, while unusually high realized volatility captures turbulent market conditions that may not coincide with a large contemporaneous decline. The 90th percentile threshold provides a transparent, real-time rule that classifies roughly the most volatile 10\% of months as stress based on information available up to $t-1$. The $-5\%$ return cutoff is chosen to represent an economically meaningful monthly drawdown. The qualitative results are similar for nearby thresholds (e.g., volatility quantiles between 0.85 and 0.95 and return cutoffs between $-3\%$ and $-7\%$).

The forecasting target is next-month stress:
\begin{equation}
Y_{t+1} \equiv S_{t+1}.
\end{equation}

\subsection{Learning MSPI}

Let $X_t\in\R^p$ denote the vector of cross-sectional fragility signals constructed from month $t$ (Table~\ref{tab:features}). MSPI is defined as the model-implied probability
\begin{equation}
MSPI_t \equiv \Pr\left(Y_{t+1}=1 \mid X_t\right) = \Lambda\left(\beta_0 + X_t^{\prime}\beta\right),
\qquad
\Lambda(z)\equiv \frac{1}{1+e^{-z}}.
\label{eq:logit}
\end{equation}

To avoid overfitting and to encourage a parsimonious mapping from predictors to stress probabilities, $(\beta_0,\beta)$ are estimated using an L1-regularized logistic regression (lasso-logit). For a training sample $\mathcal{T}$, the estimator solves:
\begin{equation}
(\hat{\beta}_0,\hat{\beta}) \in
\arg\min_{\beta_0,\beta}
\left\{
-\sum_{t\in\mathcal{T}}
\Big[
Y_{t+1}\log p_t + (1-Y_{t+1})\log(1-p_t)
\Big]
+ \lambda \|\beta\|_1
\right\},
\label{eq:lasso}
\end{equation}
where $p_t=\Lambda(\beta_0+X_t^\prime\beta)$ and $\lambda\ge 0$ is the regularization parameter. Predictors are standardized (z-scored) using information available within each training window.

A real-time expanding-window procedure is implemented. An initial training window of 120 months is used. The regularization parameter $\lambda$ is selected via time-series cross-validation within this initial window and then held fixed for the subsequent expanding-window evaluation. At each out-of-sample month $t$, the model is re-estimated using data available up to $t$, and $MSPI_t$ is computed as the predicted probability of stress in month $t+1$.

More generally, the expanding-window real-time protocol should be viewed as part of the estimator. In monitoring applications, the relevant performance object is the distribution of out-of-sample probability forecasts produced under the information constraints faced by a practitioner at the time of prediction. This motivates fixing hyperparameters after initial tuning and re-estimating model parameters each month using only data available up to that date, thereby avoiding implicit look-ahead.

\subsection{Benchmark}

To gauge incremental value, I compare MSPI to a parsimonious benchmark that uses only lagged aggregate market conditions: the previous month’s value-weighted market return
$R^{\mathrm{mkt}}_{t}$ and realized market volatility $\sigma^{\mathrm{mkt}}_{t}$.
The benchmark is a logistic regression,
\begin{equation}
p^{B}_{t}
= \Lambda\!\left(\alpha_{0} + \alpha_{1} R^{\mathrm{mkt}}_{t} + \alpha_{2} \sigma^{\mathrm{mkt}}_{t}\right),
\end{equation}
estimated with an L2-regularized penalty. Predictors are standardized within each expanding training window. The L2 penalty parameter is selected via time-series cross-validation within the initial training window and then held fixed in the subsequent expanding-window evaluation. The benchmark is evaluated in the same real-time expanding-window design as MSPI, producing one-month-ahead stress probabilities.

\subsection{Nonlinear models: random forests and gradient-boosted trees}
\label{subsec:alt_ml}

While the baseline lasso-logit mapping in equation~\eqref{eq:logit} is intentionally sparse and interpretable, stress prediction may involve nonlinear interactions among cross-sectional fragility signals. To assess the empirical importance of such nonlinearities, I compare MSPI to two standard nonparametric supervised-learning methods: random forests and gradient-boosted decision trees. Both methods use the same predictor set as MSPI (Table~\ref{tab:features}) and target the same next-month stress outcome $Y_{t+1}$.

\textbf{Random forests.}
A random forest constructs an ensemble of $B$ classification trees
$\{T_b(\cdot)\}_{b=1}^B$, each grown on a bootstrap resample of the training data. At each internal node, a split variable is selected from a random subset of predictors and the split point is chosen to minimize an impurity criterion (Gini index). For a given tree $b$, let $T_b(X_t)\in[0,1]$ denote the terminal-node stress frequency. The random-forest score is the ensemble average
\[
s^{RF}_t = \frac{1}{B}\sum_{b=1}^B T_b(X_t).
\]
Because this score is not guaranteed to be well calibrated as a probability,
I apply a real-time calibration map (Platt scaling) within each training window
to obtain the probability forecast
\[
\hat p^{RF}_t = g_{RF}(s^{RF}_t),
\]
where $g_{RF}(\cdot)$ is estimated using only information available at time $t$.

\textbf{Gradient-boosted trees.}
Gradient boosting estimates an additive model of regression trees by sequentially minimizing the Bernoulli log-likelihood loss.
Let $F_m(X)$ denote the boosted predictor after $m$ iterations, initialized at a constant $F_0$, and updated as
\[
F_m(X) = F_{m-1}(X) + \nu\, h_m(X),
\]
where $h_m(\cdot)$ is a shallow regression tree and $\nu\in(0,1]$ is a shrinkage parameter.
The loss function is
\[
\ell(Y,p) = -\big[Y\log p + (1-Y)\log(1-p)\big], \qquad p=\Lambda(F(X)).
\]
After $M$ iterations, the raw boosted score is $s^{GB}_t = F_M(X_t)$.
As with random forests, I obtain probability forecasts via real-time calibration, $\hat p^{GB}_t = g_{GB}(s^{GB}_t)$, estimated within each expanding training window.

Both nonlinear learners allow for flexible interactions among fragility signals, but do not impose a sparse or globally interpretable structure; in the present monthly monitoring setting, they therefore serve as robustness comparisons for assessing whether additional nonlinear flexibility improves out-of-sample discrimination or probability accuracy relative to the lasso-logit baseline.

All models are estimated under the same real-time expanding-window protocol. An initial training window is used to select hyperparameters via time-series cross-validation (forward-chaining splits), after which the chosen hyperparameters are held fixed for the subsequent expanding-window evaluation. At each out-of-sample month $t$, the model is re-estimated using information available up to $t$ and produces a one-month-ahead probability forecast of stress. Randomization is controlled via fixed seeds to ensure reproducibility.

The goal of these nonlinear models is not to replace the interpretable baseline, but to provide a modern machine-learning horse race that clarifies whether additional complexity improves out-of-sample discrimination and probability reliability in this real-time stress-measurement setting. Because nonparametric learners can produce poorly calibrated probability outputs, I also report calibrated variants using Platt scaling estimated in real time within each training window; this calibration step targets probability accuracy (Brier score, log loss, and ECE) rather than ranking metrics.

\begin{tcolorbox}[
  enhanced,
  breakable,
  title=\textbf{Box 2. Real-Time Construction of MSPI},
  colback=white,
  colframe=black!80,
  fonttitle=\bfseries
]

At each month $t$, MSPI is updated using only information available through month $t$:

\begin{enumerate}\itemsep2pt
\item \textbf{Compute predictors.}
Using daily CRSP data in month $t$, compute the cross-sectional fragility signals
(dispersion, tail shares, higher moments, trading-intensity proxies) and aggregate them
to monthly features $X_t$.

\item \textbf{Update the stress label history.}
Classify past months as stress/non-stress using the fixed return cutoff and the expanding,
real-time volatility quantile computed using data available at the time.

\item \textbf{Re-estimate the mapping.}
Refit the lasso-logit on an expanding window using $\{(X_\tau, Y_{\tau+1})\}_{\tau \le t}$,
with standardization computed within the training window.

\item \textbf{Produce the forecast.}
Output $\text{MSPI}_t = \Pr(Y_{t+1}=1 \mid X_t)$ as the one-month-ahead stress probability.

\item \textbf{(If using nonlinear benchmarks) Calibrate in real time.}
Estimate a calibration map within the training window and report probabilities on the
calibrated scale.
\end{enumerate}

This discipline (expanding window, no look-ahead, reproducible updating) is essential when
the object is a monitoring input rather than an ex post index.

\end{tcolorbox}

\section{Results}\label{sec:results}

This section summarizes the out-of-sample behavior of MSPI and compares it to the benchmark and nonlinear alternatives along three dimensions: time-series dynamics, discrimination performance, and probability accuracy (calibration).

\subsection{Time-series behavior}

\begin{figure*}[!htbp]
\centering
\includegraphics[width=\linewidth]{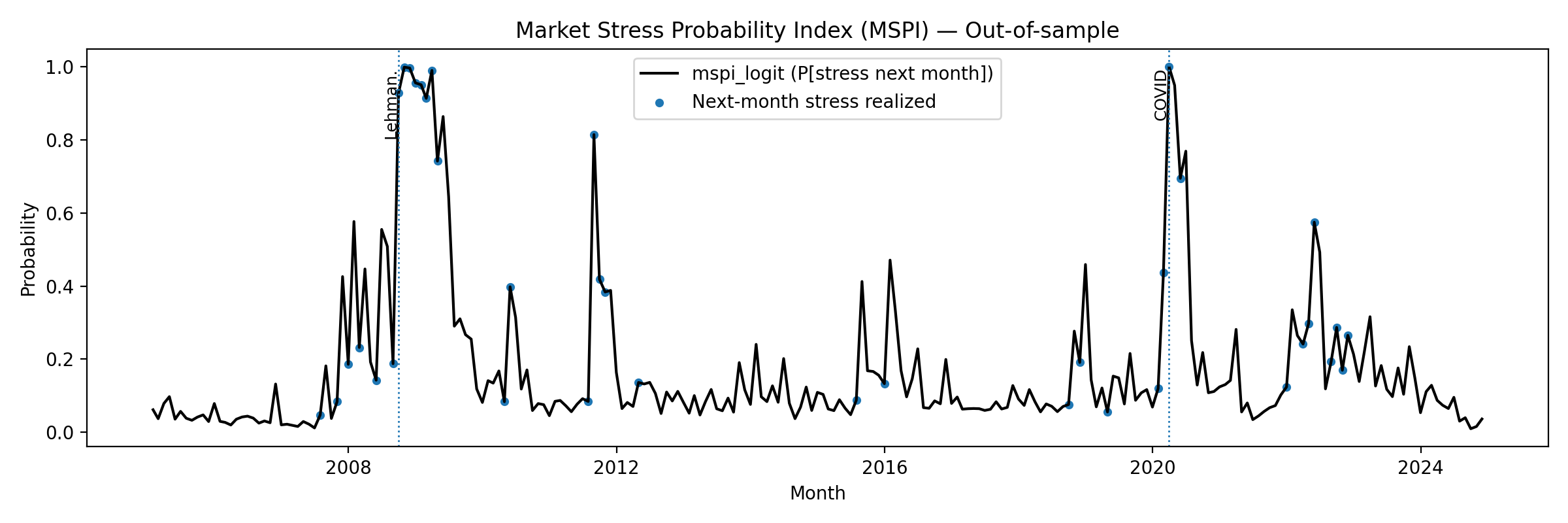}
\caption{MSPI out-of-sample. The solid line plots the predicted probability that the \emph{next} month is a stress month. Dots mark months for which stress is realized in the subsequent month under the definition in equation~\eqref{eq:stress_def}.}
\label{fig:mspi_ts}
\end{figure*}

Figure~\ref{fig:mspi_ts} plots MSPI over the out-of-sample period (2005--2024). The index rises ahead of major stress episodes, with prominent spikes during the 2008--2009 crisis and the 2020 COVID shock. The figure also shows that elevated MSPI tends to cluster, consistent with persistence in stressful market states.

\begin{figure*}[!htbp]
\centering
\includegraphics[width=\linewidth]{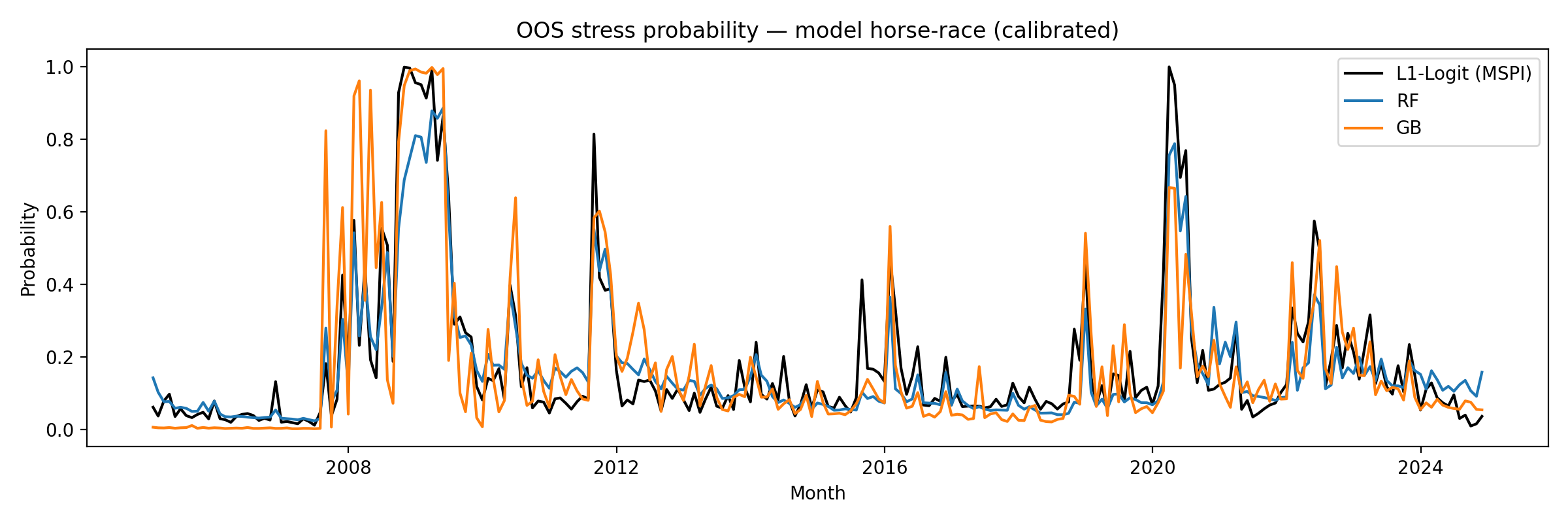}
\caption{Out-of-sample stress probability - model horse race (calibrated). MSPI (black) is the L1-logit probability of next-month stress. Random forest (blue) and gradient boosting (orange) are estimated under the same expanding-window design and Platt-calibrated in real time within each training window to place forecasts on a comparable probability scale.}
\label{fig:horse_race_ts}
\end{figure*}

To benchmark the time-series behavior of the probability signal itself, Figure~\ref{fig:horse_race_ts} overlays MSPI with two flexible nonlinear classifiers (random forests and gradient-boosted trees) estimated under the same real-time expanding-window protocol. Because the object of interest is a \emph{probability-of-stress} measure, I plot RF and GB on a calibrated probability scale using Platt scaling estimated in real time within each training window.

The figure highlights two patterns. First, all three models assign elevated stress probabilities ahead of major stress episodes, most notably around the 2008--2009 crisis and the 2020 COVID shock, indicating that the cross-sectional fragility signals contain robust forward-looking information about stress regimes. Second, the nonlinear learners produce a more jagged and episodic probability path—especially in pre-crisis and transition periods—whereas MSPI delivers a comparatively smoother and more stable probability signal. This stability is valuable for monitoring and applied work: it supports interpreting MSPI as a disciplined, real-time state variable rather than a purely opportunistic classification score. Consistent with this interpretation, the probability-accuracy metrics reported in Table~\ref{tab:oos_metrics} (Brier score, log loss, and calibration diagnostics) favor the sparse logit specification even when flexible benchmarks are tuned and calibrated under the same protocol.

I next quantify these comparisons using rank-based discrimination metrics (ROC and precision--recall) computed from raw scores, and probability-accuracy metrics computed on the calibrated probability scale.

\begin{figure}[!htbp]
\centering
\includegraphics[width=0.75\linewidth]{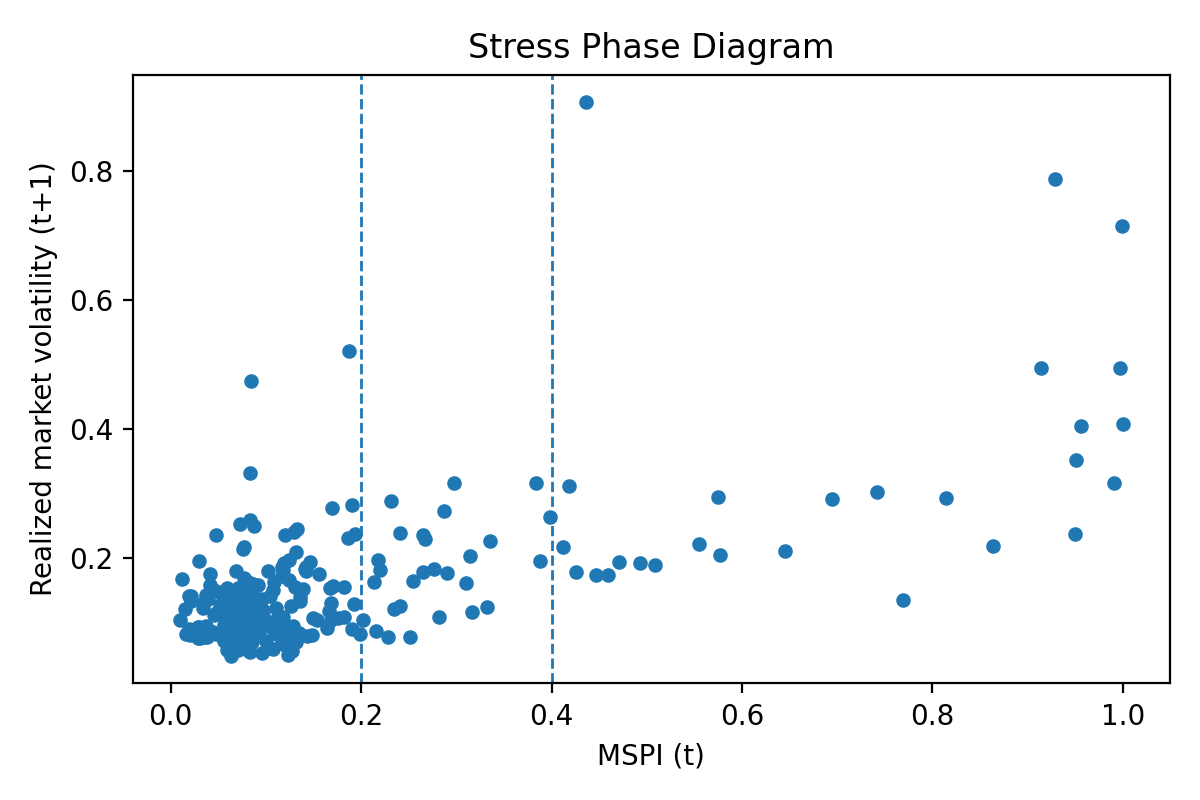}
\caption{Stress phase diagram. Each point plots MSPI$_t$ against next-month realized market volatility $\sigma^{mkt}_{t+1}$. The strong positive association supports interpreting MSPI as a forward-looking risk state variable.}
\label{fig:stress_phase}
\end{figure}

Figure~\ref{fig:stress_phase} provides a complementary visualization by relating current stress probabilities to next-month realized market volatility. The positive association is consistent with interpreting MSPI as a forward-looking risk state variable rather than a purely contemporaneous stress indicator.

\subsection{Out-of-sample performance}

Table~\ref{tab:oos_metrics} reports out-of-sample performance for MSPI, the benchmark, and two nonlinear models (random forests and gradient-boosted trees). MSPI achieves an AUC of 0.800 and a PR-AUC of 0.538, improving upon the benchmark (AUC 0.752; PR-AUC 0.444). In a horse race, gradient boosting attains an AUC of 0.756 and a PR-AUC of 0.481, while the random forest is weaker in discrimination (AUC 0.727; PR-AUC 0.447). Importantly, MSPI also delivers strong probability accuracy: it attains the lowest Brier score (0.106) and log loss (0.352) among the evaluated models, consistent with the goal of constructing a reliable probability-of-stress measure rather than a purely ranking-based classifier.

Because MSPI is interpreted on a probability scale, calibration is a first-order design objective. MSPI’s mean predicted probability (0.180) is close to the realized stress event rate (0.159), and its expected calibration error (ECE 0.062) improves upon the benchmark (ECE 0.080). The nonlinear results illustrate a common discrimination--calibration trade-off: while gradient boosting is competitive in AUC, it exhibits larger probability errors (Brier 0.114; log loss 0.396; ECE 0.071) than MSPI.

\begin{table*}[t]
\centering
\caption{Out-of-sample performance (2005--2024): discrimination and probability accuracy.}
\label{tab:oos_metrics}
\small
\setlength{\tabcolsep}{6pt}
\renewcommand{\arraystretch}{1.1}

\input{tables/oos_metrics_all_calibrated.tex}

\vspace{0.25em}
\footnotesize\emph{Notes: AUC and PR-AUC are computed from raw (uncalibrated) model scores. Brier score, log loss, ECE, and mean predicted probability are computed from probability forecasts; for the nonlinear learners (RF and GB), probabilities are Platt-calibrated in real time within each training window, while MSPI (L1-logit) and the ridge-logit benchmark use their native logistic probabilities. ECE is computed using 10 equal-mass (quantile) bins. Lower Brier score, log loss, and ECE indicate better probability accuracy.} 
\end{table*}

\begin{figure}[!htbp]
\centering
\includegraphics[width=0.75\linewidth]{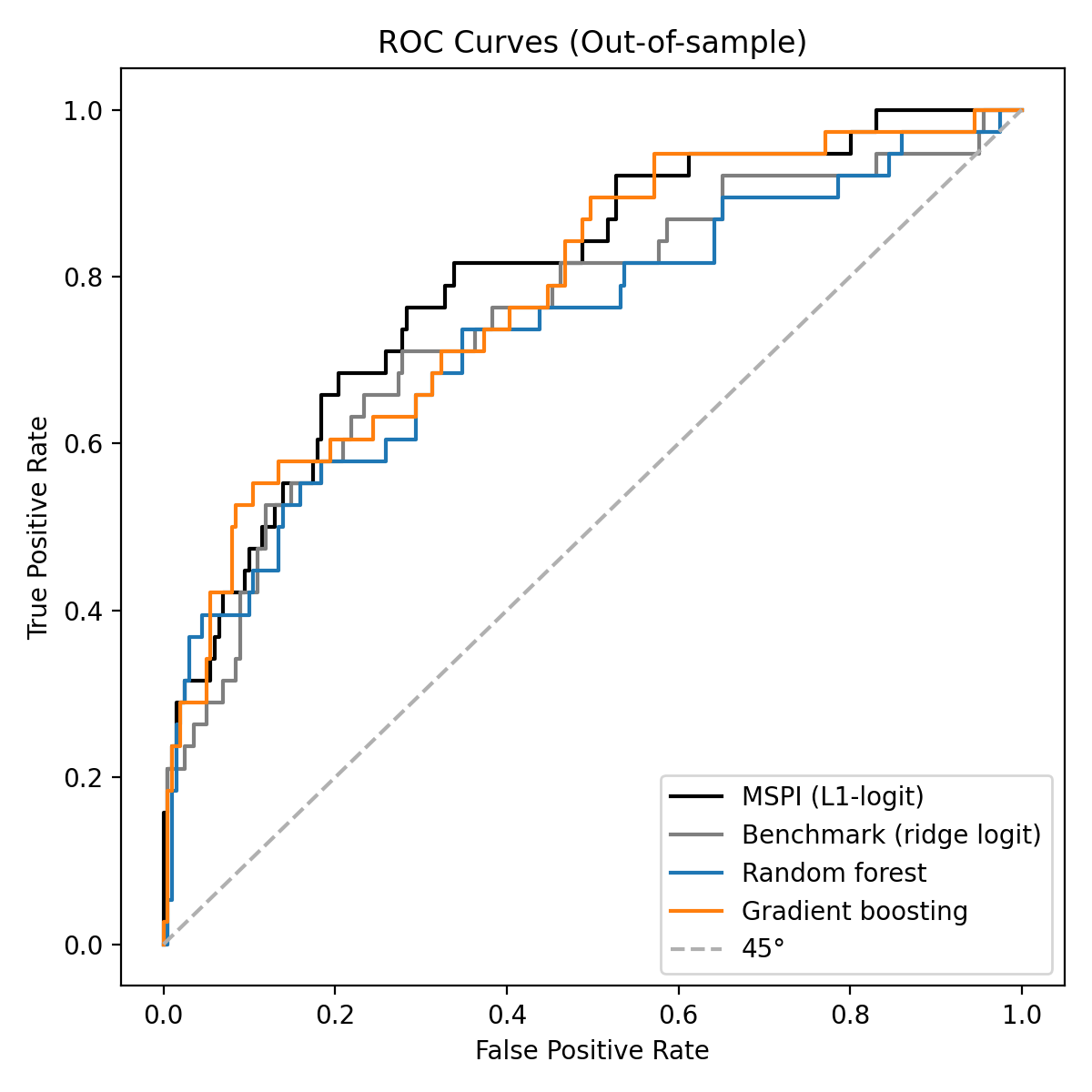}
\caption{ROC curves out-of-sample (raw scores). Curves are computed from uncalibrated model scores to measure discrimination; the dashed $45^\circ$ line corresponds to random classification.}
\label{fig:roc}
\end{figure}

\begin{figure}[!htbp]
\centering
\includegraphics[width=0.75\linewidth]{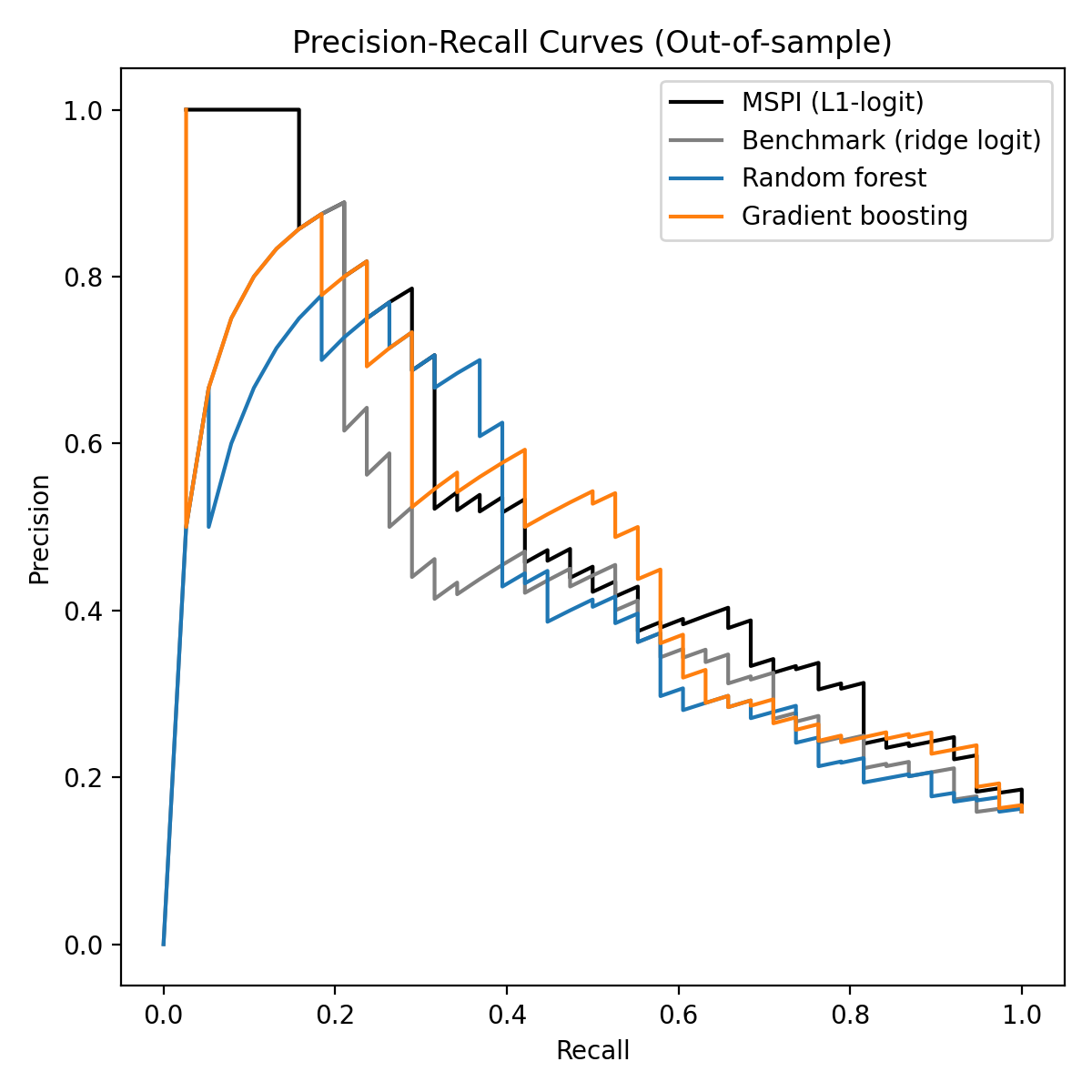}
\caption{Precision--recall curves out-of-sample (raw scores). Curves are computed from uncalibrated model scores; precision--recall is informative here because stress months are relatively infrequent (event rate 15.9\%).}
\label{fig:pr}
\end{figure}

\begin{figure}[!htbp]
\centering
\includegraphics[width=0.75\linewidth]{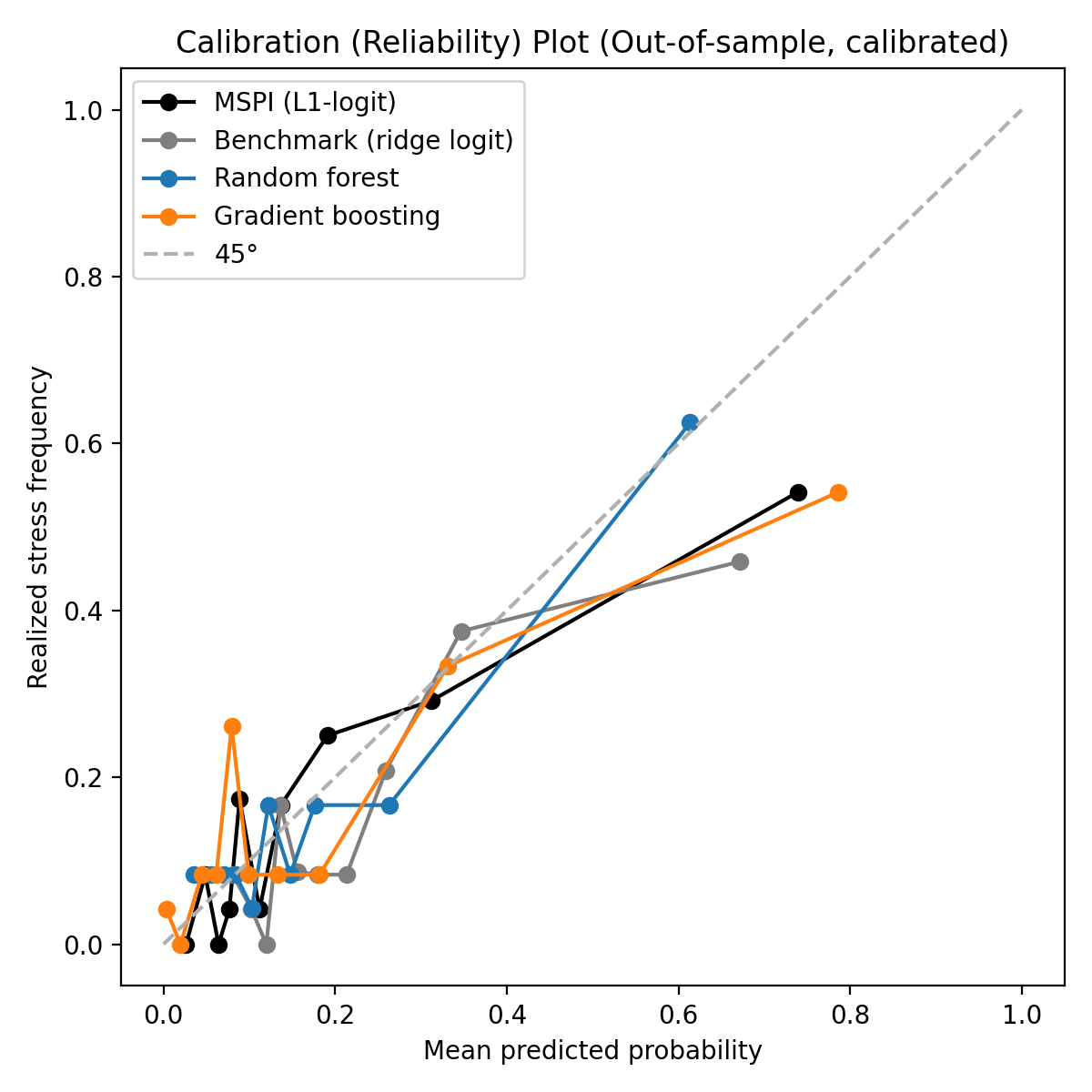}
\caption{Calibration curves out-of-sample (probability forecasts). The figure plots realized stress frequencies against predicted probabilities in equal-mass bins. Random forest and gradient boosting probabilities are Platt-calibrated in real time within each training window; MSPI and the ridge-logit benchmark are direct logistic probability forecasts.}
\label{fig:calibration}
\end{figure}

Figure~\ref{fig:roc} visualizes the same comparison using ROC curves. Across false-positive rates, MSPI delivers higher true-positive rates than the benchmark, consistent with the AUC improvement. 
Figure~\ref{fig:pr} complements the ROC evidence by focusing on precision--recall trade-offs in an imbalanced-outcome setting, where stress months are relatively infrequent.
Figure~\ref{fig:calibration} shows that MSPI’s predicted probabilities closely track realized stress frequencies across probability bins, consistent with its strong calibration metrics reported in Table~\ref{tab:oos_metrics}.

Beyond ranking performance, MSPI is designed as a probability measure and therefore places weight on calibration and real-time implementability. The improvements in Brier score and log loss indicate that MSPI delivers more accurate probability assessments of next-month stress, not only better classification. The construction is also operational: both predictors and the stress label are formed using only information available at the time of prediction, and estimation is carried out in an expanding-window design. This real-time discipline is essential if MSPI is to be used as a monitoring input rather than as an ex post descriptive index.

\subsection{Bootstrap inference for performance differences}

To quantify sampling uncertainty in out-of-sample performance and account for serial dependence, I implement a block bootstrap over months (12-month blocks; 2{,}000 replications). Table~\ref{tab:bootstrap} reports differences relative to the benchmark for AUC, PR-AUC, Brier score, and log loss. The point estimates favor MSPI across all four metrics, but the confidence intervals are conservative and do not reject equality at conventional levels in this sample length, consistent with limited power in monthly out-of-sample evaluation. Because differences are computed within each bootstrap resample and then averaged, the reported $\Delta$ statistics may differ slightly from the corresponding point-estimate differences in Table~\ref{tab:oos_metrics}. Finite-sample distortions of asymptotic inference are well known in econometrics \citep{WINDMEIJER200525}.

\begin{table*}[!htbp]
\centering
\caption{Block-bootstrap inference for out-of-sample performance differences relative to the benchmark.}
\label{tab:bootstrap}
\small
\setlength{\tabcolsep}{6pt}
\renewcommand{\arraystretch}{1.1}

\input{tables/bootstrap_diffs_calibrated}

\vspace{0.25em}
\footnotesize\emph{Notes:} $\Delta$ denotes the mean difference in the performance metric between the indicated model and the benchmark, computed across block-bootstrap resamples of the out-of-sample period. As a result, the reported $\Delta$ values need not coincide exactly with the single point-estimate differences implied by Table~\ref{tab:oos_metrics}. For AUC and PR--AUC, positive values indicate improvement relative to the benchmark; for Brier score, log loss, and ECE, negative values indicate improvement.

\end{table*}

\section{Economic Interpretation and Algorithmic Monitoring}\label{sec:econ}

The primary purpose of MSPI is measurement: it provides a forward-looking, real-time probability that the U.S. equity market will enter a high-stress regime over the next month. This object is best viewed through the lens of \emph{financial econometrics in the age of AI}: as trading, liquidity provision, and execution become increasingly electronic and algorithmically mediated, it is natural that monitoring and measurement also shift toward algorithmic systems that can aggregate high-dimensional market information into transparent, reproducible state variables
\citep{HendershottJonesMenkveld2011,KirilenkoKyleSamadiTuzun2017,CHABOUD2014}.
In this setting, the key requirement is not maximal model complexity, but reliable probabilistic measurement under real-time constraints. MSPI is designed accordingly: it is computed from widely available equity-market data, estimated in an expanding-window real-time protocol, and evaluated using probability accuracy and calibration diagnostics that are appropriate for decision-relevant probability forecasts
\citep{Brier1950,GneitingRaftery2007}.

\subsection{Economic value on a probability scale}

A probability-of-stress measure is useful precisely because it can be mapped into transparent decision thresholds while remaining comparable across samples and estimation windows. This probability scale is also what makes calibration first-order: poorly calibrated classifiers may rank stress well, yet deliver misleading probability levels. Proper scoring rules such as the Brier score and log loss therefore provide a coherent way to evaluate whether probability forecasts are informative \emph{and} reliable \citep{Brier1950,GneitingRaftery2007}. For the nonlinear benchmarks, explicit probability calibration (e.g., Platt scaling) can improve reliability and is a standard component of a
probabilistic forecasting pipeline \citep{Platt1999,ZadroznyElkan2002}.

\subsection{Realized outcomes across MSPI bins}

\begin{table*}[!htbp]
\centering
\caption{Realized outcomes by MSPI bins (out-of-sample).}
\label{tab:binned}
\small
\setlength{\tabcolsep}{7pt}
\renewcommand{\arraystretch}{1.1}

\input{tables/binned_outcomes_raw}

\end{table*}

A simple interpretation of MSPI is obtained by examining how realized outcomes vary across predicted probability levels. Table~\ref{tab:binned} bins months by MSPI and reports (i) the realized frequency of stress in the subsequent month, (ii) subsequent realized
market volatility, and (iii) subsequent market returns.

Two patterns stand out. First, the realized next-month stress frequency rises sharply across MSPI bins, indicating that the index carries
meaningful forward-looking information about regime-like stress realizations. Second, higher MSPI is associated with substantially higher subsequent realized volatility, consistent with interpreting MSPI as a near-term risk state variable. This interpretation is closely related to the idea that tail-like risks can be inferred from equity-market data, including from the cross-section of individual stock returns \citep{KellyJiang2014}.

The return patterns in Table~\ref{tab:binned} should be interpreted cautiously. At the monthly horizon, realized returns in stress-prone periods can reflect both downside tail events and subsequent rebounds, and mean returns within bins are estimated with noise. The more robust
economic content of MSPI is therefore its ability to summarize near-term stress risk and volatility dynamics on a calibrated probability scale.

\subsection{Predictive interpretation for volatility and downside-tail risk}

A natural finance-native application is one-step-ahead prediction of risk outcomes. To formalize the interpretation of MSPI as a forward-looking risk state variable, I illustrate how it can be used to predict next-month realized market volatility. Realized volatility constructed from high-frequency return observations is a standard empirical measure of time-varying risk \citep{AndersenBollerslevDieboldLabys2003,BarndorffNielsenShephard2002}.
A simple predictive regression specification is
\begin{equation}
\sigma^{mkt}_{t+1} = \alpha + \gamma \, MSPI_t + \phi' Z_t + \eta_{t+1},
\label{eq:pred_vol}
\end{equation}
where $Z_t$ may include parsimonious lagged controls such as $R^{mkt}_t$ and $\sigma^{mkt}_t$. Because MSPI is constructed from cross-sectional fragility signals rather than only lagged aggregates, equation~\eqref{eq:pred_vol} provides a transparent way to assess incremental predictive content for near-term risk dynamics.

A complementary tail-oriented outcome is a downside indicator at the monthly horizon,
\begin{equation}
Crash_{t+1}(c) \equiv \indic\{R^{mkt}_{t+1} \le c\},
\end{equation}
which can be studied in a linear probability or logistic specification. These predictive specifications emphasize the intended role of MSPI: a disciplined probability state variable that can forecast financially meaningful states (volatility and downside-tail realizations), rather than a structural shock with causal interpretation.

\subsection{From measurement to stress-risk innovations}\label{subsec:innov}

A limitation of purely descriptive stress indices is that they are often less convenient for empirical designs that require a shock-like object. I therefore construct a stress-risk innovation by extracting the component of MSPI that is orthogonal to lagged information. Concretely, I project $MSPI_t$
on lagged MSPI and lagged market controls,
\begin{equation}
MSPI_t = \delta_0 + \delta_1 MSPI_{t-1} + \Delta' Z_{t-1} + u_t,
\label{eq:innov}
\end{equation}
where $Z_{t-1}$ includes lagged market return and realized volatility. I interpret the residual $u_t$ as ``news'' about near-term stress risk: it is, by construction, unpredictable from the chosen lagged information set. This innovation is convenient for dynamic regressions and for studying how changes in near-term stress risk co-move with subsequent financial outcomes.

\subsection{Dynamic financial-state associations: a local-projection template}\label{subsec:lp}

To summarize dynamic associations in an interpretable way, I use local projections to relate a stress-risk innovation to subsequent financial outcomes. For horizons $h=0,1,\dots,H$, I estimate
\begin{equation}
y_{t+h} = a_h + b_h u_t + \Gamma_h' W_{t-1} + \varepsilon_{t+h},
\label{eq:lp}
\end{equation}
where $u_t$ is the stress-risk innovation defined in equation~\eqref{eq:innov} and $W_{t-1}$ contains lagged controls. Natural choices for $y_{t+h}$ in this context are finance-native state variables computed from the same CRSP universe, including realized market volatility, downside-tail indicators, and next-month measures of cross-sectional fragility (e.g., dispersion or extreme-return shares). 

Standard errors can be computed using HAC estimators that account for serial correlation induced by overlapping horizons. The coefficients $\{b_h\}_{h=0}^H$ summarize how financial states tend to evolve following a shock-like increase in near-term stress risk. These estimates are best interpreted as dynamic correlations conditional on the chosen controls; stronger causal interpretation would require additional identifying assumptions.

\subsection{Algorithmic monitoring of algorithmic markets}

The motivation for MSPI is closely tied to market structure. Modern equity markets are increasingly shaped by automated execution, electronic liquidity provision, and algorithmic decision rules \citep{HendershottJonesMenkveld2011,KirilenkoKyleSamadiTuzun2017}. Stress can therefore emerge from rapidly shifting cross-sectional dynamics and nonlinear amplification mechanisms that interact with market design. For instance, circuit breakers can alter behaviour as the market approaches a halt, generating magnet effects and shifts in return skewness and trading activity \citep{CHEN2024}. More broadly, informational frictions can create ``black holes'' in which information aggregation fails precisely when screening and learning are most valuable \citep{AXELSON2023}, and systemic risk can reflect common exposures that build endogenously over the cycle \citep{KOPYTOV2023}.

In such an environment, a practical monitoring system should satisfy three design principles. First, it should be implementable and backtestable in real time. Second, it should be robust and transparent enough to be used as an input in governance and monitoring workflows, where model risk and interpretability matter. Third, because monitoring decisions are often threshold-based, the output should be a calibrated probability rather than an arbitrary score \citep{Brier1950,GneitingRaftery2007}. These principles favor parsimonious probabilistic models with explicit evaluation of probability accuracy and calibration, and they motivate the
paper's emphasis on lasso-logit as a disciplined baseline \citep{Tibshirani1996}, alongside nonlinear benchmarks that clarify the incremental value of additional flexibility \citep{Breiman2001,Friedman2001}.

Finally, the monitoring interpretation aligns with the broader evidence that modern financial decision-making increasingly combines automated tools with human oversight. Algorithmic decision aids often complement rather than fully replace human judgement in investment-related settings \citep{Bianchi2024,Cao2024}, and machine-learning benchmarks can serve as standards for diagnosing systematic expectation errors
\citep{vanBinsbergen2023}. In this spirit, MSPI is best interpreted as an interpretable monitoring input: a reproducible probability-of-stress signal that can be combined with domain expertise and additional information sources. Potential uses include studying the transmission of stress across assets and incorporating MSPI as a control or state variable in empirical models of volatility, tail risk, and risk premia; practitioner-facing threshold rules are a natural operational interpretation of the probability scale.

\section{Conclusion}\label{sec:conclusion}

This paper proposes an equity-only measure of near-term market stress risk: the Market Stress Probability Index (MSPI). MSPI is constructed from a set of cross-sectional fragility signals and estimated using an L1-regularized logistic model in a real-time expanding-window design. Out of sample, MSPI tracks major stress episodes and delivers improved accuracy and discrimination relative to a parsimonious benchmark based only on lagged market return and realized volatility. By design, MSPI is transparent, easily updated, and implementable using only CRSP data, making it suitable as a real-time monitoring input rather than an ex post descriptive index. The probability scale facilitates direct use in empirical macro-finance and financial econometrics, where MSPI can be embedded as a forward-looking stress state variable. Future work can explore alternative stress definitions, richer economic applications, and extensions to multi-asset settings.






\bibliographystyle{oup-abbrvnat}
\bibliography{reference}







\end{document}

%% file: tables/oos_metrics_all_calibrated.tex
\begin{tabular}{lcccccccc}
\toprule
Model & AUC & PR--AUC & Brier & Log loss & ECE & Mean prob. & Event rate & $N$ \\
\midrule
MSPI (L1-logit) & 0.800 & 0.538 & 0.106 & 0.352 & 0.062 & 0.180 & 0.159 & 239 \\
Benchmark (ridge logit) & 0.752 & 0.444 & 0.116 & 0.400 & 0.080 & 0.227 & 0.159 & 239 \\
Random forest & 0.727 & 0.447 & 0.111 & 0.377 & 0.037 & 0.167 & 0.159 & 239 \\
Gradient boosting & 0.756 & 0.481 & 0.114 & 0.396 & 0.071 & 0.174 & 0.159 & 239 \\
\bottomrule
\end{tabular}

%% file: tables/bootstrap_diffs_calibrated.tex
\begin{tabular}{llcccc}
\toprule
Metric & Model & $\Delta$ vs.\ bench & 95\% CI (lo) & 95\% CI (hi) & $p$-value \\
\midrule
AUC & MSPI (L1-logit) & 0.044 & -0.049 & 0.153 & 0.376 \\
AUC & Random forest & -0.016 & -0.119 & 0.079 & 0.746 \\
AUC & Gradient boosting & 0.024 & -0.052 & 0.109 & 0.552 \\
\midrule
PR--AUC & MSPI (L1-logit) & 0.079 & -0.046 & 0.208 & 0.240 \\
PR--AUC & Random forest & 0.001 & -0.125 & 0.122 & 0.992 \\
PR--AUC & Gradient boosting & 0.046 & -0.055 & 0.134 & 0.353 \\
\midrule
Brier & MSPI (L1-logit) & -0.009 & -0.025 & 0.009 & 0.336 \\
Brier & Random forest & -0.004 & -0.021 & 0.011 & 0.595 \\
Brier & Gradient boosting & -0.001 & -0.020 & 0.023 & 0.863 \\
\midrule
Log loss & MSPI (L1-logit) & -0.044 & -0.112 & 0.016 & 0.157 \\
Log loss & Random forest & -0.021 & -0.087 & 0.041 & 0.530 \\
Log loss & Gradient boosting & 0.002 & -0.084 & 0.104 & 0.983 \\
\midrule
ECE & MSPI (L1-logit) & -0.015 & -0.056 & 0.041 & 0.518 \\
ECE & Random forest & -0.023 & -0.061 & 0.027 & 0.302 \\
ECE & Gradient boosting & 0.001 & -0.050 & 0.059 & 0.983 \\
\bottomrule
\end{tabular}

%% file: tables/binned_outcomes_raw.tex
\begin{tabular}{lccccc}
\toprule
MSPI bin & $N$ & Mean prob. & Next-mo.\ stress rate & Next-mo.\ mkt.\ vol. & Next-mo.\ mkt.\ ret. \\
\midrule
$[0.00,0.05)$ & 37 & 0.032 & 0.027 & 0.116 & 0.010 \\
$[0.05,0.10)$ & 83 & 0.074 & 0.072 & 0.122 & 0.007 \\
$[0.10,0.20)$ & 64 & 0.139 & 0.156 & 0.144 & 0.008 \\
$[0.20,0.40)$ & 28 & 0.277 & 0.250 & 0.184 & 0.018 \\
$[0.40,1.00]$ & 27 & 0.703 & 0.519 & 0.331 & 0.006 \\
\bottomrule
\end{tabular}